\documentclass[prb,superscriptaddress,floatfix]{revtex4}
\usepackage{color,epsfig}
\usepackage{bm}
\usepackage{amsmath,amsfonts,amssymb,bm}
\usepackage{natbib}

\begin{document}
\title{The renormalized superperturbation theory (SPT) approach to the Anderson model in and out of equilibrium}

\author{Enrique Mu\~{n}oz}
\address{Facultad de F\'isica, Pontificia Universidad Cat\'olica de Chile, Vicu\~{n}a Mackenna 4860, Santiago, Chile}

\author{Farzaneh Zamani}
\address{Max Planck Institute for the Physics of Complex Systems, 01187 Dresden, Germany}
\address{Max Planck Institute for Chemical Physics of Solids, 01187 Dresden, Germany}

\author{Lukas Merker}
\address{Peter Gr\"unberg Institut and Institute for Advanced Simulation, Research Centre J\"ulich, D-52425 J\"ulich, Germany}

\author{Theo Costi}
\address{Peter Gr\"unberg Institut and Institute for Advanced Simulation, Research Centre J\"ulich, D-52425 J\"ulich, Germany}

\author{Stefan Kirchner}
\address{Center for Correlated Matter, Zhejiang, University, Hangzhou, China}

\begin{abstract}
The properties of current-carrying steady states of strongly correlated systems away from the linear-response regime are of topical interest.
In this article, we review the renormalized perturbation theory, or renormalized SPT of reference \cite{Munoz.13} for the Anderson model. We present an extension  to higher orders and compare the higher-order results with NRG calculations. Finally, we  elucidate the role of Ward identities in calculating out-of-equilibrium properties and address claims made in the literature.
\end{abstract}

\maketitle

\section{Introduction}

Quantum dots as well as other artificial nanostructures, such as single-molecules attached to conducting leads, constitute very versatile devices, since their characteristic energy scales such as Fermi level,
charging energy, level spacing can be tuned over a wide range \cite{GoldhaberGordon.98,Cronenwett.98,Potok.07}. The high degree of characterization of these systems has made it possible to investigate the properties of strongly-correlated electrons away from equilibrium in a well-defined setting~\cite{Grobis.08,Scott.09}.
In particular, the
low-temperature, low-bias differential conductance characteristics in these nanostructures often display a nearly universal conductance enhancement due to  Kondo
screening of the local, {\it i.e.}, localized on the quantum dot,  dynamic degree of freedom by the lead electrons \cite{Hewson}. 
It is well established, that 
in equilibrium, i.e. within the linear response regime, the Kondo effect leads to an enhancement of the differential conductance
$G = dI/dV|_{V=0}$ on the order of twice the quantum of conductance $2e^{2}/h$ in the limit $T\rightarrow 0$. 
These experimental results 
are usually interpreted within the context of the single-level Anderson model~\cite{Glazman.88,Ng.88}
\begin{equation}
\label{Andersonmodel}
H_{\mbox{\tiny AM}}=\sum_{\lambda=L/R} H_0^\lambda(c_\lambda^\dagger,c^{}_{\lambda})+\sum_{\sigma=\pm} \epsilon_d d^\dagger_\sigma d^{}_\sigma+Ud^\dagger_{+}d^{}_{+}d^\dagger_{-}d^{}_{-}+\sum_{\lambda}\sum_{\sigma}\big ( V d^\dagger_{\sigma}c_{0\sigma\lambda}+ \mbox{h.c.}\big)
\end{equation}
as the effective low-energy model. In Eq.~(\ref{Andersonmodel}),  $H_0^\lambda$ is a free electron Hamiltonian given in terms of $c_\lambda^\dagger$ and $ c^{}_{\lambda}$ which describes the excitations of lead $\lambda$, $c^{\dagger}_{0\sigma\lambda}$ is the corresponding creation operator of lead $\lambda$ at the quantum dot site $\mathbf{r}=0$. $V$ denotes the coupling strength between the lead states and the quantum dot states, described by $d_\sigma$ and $U$ is the charging energy of the dot.

Correspondingly, a large
class of conductance experiments has been accurately fitted by an expression of the form \cite{Grobis.08,Scott.09,Kaminski.00,Doyon.06,Schiller.95} 
\begin{eqnarray}
\frac{G(0,0) - G(T,V)}{c_{T} G(0,0)} = \left(\frac{T}{T_{K}} \right)^{2} + \alpha\left(\frac{eV}{k_{B}T_{K}}  \right)^{2} -\gamma c_{T}\left(\frac{eVT}{k_{B}T_{K}^{2}} \right)^{2},
\label{eqIntro1}
\end{eqnarray}
where $T_K$ is some characteristic energy scale which is obtained from the fit to the experimental results. When casting the low-temperature non-linear conductance of the Anderson model, Eq.~(\ref{Andersonmodel}) into the form (\ref{eqIntro1}),  $T_K$ becomes the Kondo temperature.
The coefficients $\alpha$ and $\gamma$ vary across different experimental systems  \cite{Grobis.08,Scott.09}. Already addressing how  $\alpha$ and $\gamma$ depend on the parameters of $H_{\mbox{\tiny AM}}$ is a difficult task as it requires a proper description of the low-energy excitations of the Anderson model in a non-thermal steady state. While there is a rather complete picture of the equilibrium behavior of Eq.~(\ref{Andersonmodel}), such an understanding for the out-of-equilibrium properties is still lacking. This is largely due to the lack of reliable methods that can tackle the strongly correlated nature of the non-thermal steady state associated with $H_{\mbox{\tiny AM}}$.

The equilibrium properties of the Anderson model have been studied extensively with the help of {\it e.g.} numerical renormalization group methods (NRG)~\cite{Costi.94c}, pseudoparticle~\cite{Kirchner.04} and Monte Carlo methods. Analytical methods
have also been developed to study this problem, such as the bare perturbation
theory of Yamada and Yosida \cite{Yosida.70,Yamada.75,Yamada.79}. An exact analytical solution has been obtained via Bethe Ansatz \cite{Andrei.83,Tsvelic.83,Zlatic.83}. 
Unfortunately, fully reliable generalizations of these methods to the 
out-of-equilibrium
problem are, at least at present, not available. However, a number of perturbative renormalization group (RG) methods
have been proposed for non-equilibrium systems \cite{Rosch_03,Rosch_05}, including functional
perturbative RG \cite{Jakobs_07}.
Among the perturbative methods available, the renormalized perturbation theory (RPT) developed
by Hewson \cite{Hewson,Hewson.93,Hewson.01,Hewson.11} provides a framework that can in principle be extended out of equilibrium, for instance by relying on exact Ward identities \cite{Oguri.05,Hewson.05}.
The RPT provides an accurate description of the equilibrium Fermi-liquid regime
in terms of quasi-particle excitations, characterized by effective interaction parameters \cite{Hewson,Hewson.93,Hewson.01}. 

Motivated by the above-mentioned experiments, following Refs.~\cite{Hewson.01,Oguri.05},
and in the spirit of Ref.~\cite{Hafermann.09}, 
we recently developed the SPT ~ \cite{Munoz.13} - a renormalized superperturbation theory in terms of dual fermions~\cite{Rubtsov.08,Jung.11} on the Keldysh contour.
Using this method, we studied steady-state non-equilibrium transport in the single-level Anderson model beyond particle-hole symmetry,
by constructing a perturbative scheme based on a particle-hole symmetric reference system \cite{Munoz.13}. 
We constructed the reference system non-equilibrium self-energy at finite bias by extending
Oguri's Ward identity approach \cite{Oguri.05,suppl}, which has the advantage of providing a current conserving expansion by construction \cite{Baym.62,Munoz.13,suppl}. 
The dual fermion, an auxiliary fermionic degree of freedom, is used to construct a systematic expansion around the interacting reference system~\cite{Rubtsov.08}.

Within the SPT, we have studied the role of level asymmetry (gate voltage) and local Coulomb repulsion (charging energy) on the non-linear conductance of the single-level Anderson model in the steady-state regime \cite{Munoz.13,Merker.13}. 
A comparison of our analytic results for the linear response transport coefficients with NRG calculations demonstrated an excellent quantitative agreement, even at relatively large level asymmetry~\cite{Merker.13}. Moreover, our results have recently provided a theoretical framework to describe and interpret magneto-transport experiments in single-molecule transistors \cite{Scott.13}.

In this article, we review the SPT and extend the calculation to higher order (in the renormalized interaction) than reported before. In section 5, we compare the higher order calculation with NRG calculations for the linear-response transport coefficients. In the appendix, we  exemplify the proof of Oguri's Ward identity~\cite{Oguri.01} and address claims made in the literature regarding the validity of this identity.

\section{Path-Integral representation on the Keldysh contour}

We consider the single-level Anderson model, Eq.~(\ref{Andersonmodel}), which can be represented  by a non-equilibrium functional on the Keldysh
contour~\cite{Munoz.13},
\begin{eqnarray}
Z = \int\mathcal{D}[\hat{\psi}^{\dagger},\hat{\psi}]\,\mathcal{D}[\hat{\Phi}^{\dagger},\hat{\Phi}]
e^{iS[\hat{\psi}^{\dagger},\hat{\psi},\hat{\Phi}^{\dagger},\hat{\Phi}]},
\label{eqAC1}
\end{eqnarray}
where the action in the Schwinger-Keldysh matrix representation is
\begin{eqnarray}
S[\hat{\psi}^{\dagger},\hat{\psi},\hat{\Phi}^{\dagger},\hat{\Phi}]
&=&\int_{-\infty}^{+\infty}dt\left\{\sum_{k,\lambda,\sigma}\hat{\psi}^{\dagger}_{k\lambda\sigma}(t)
\left(i\partial_{t} - \epsilon_{k\lambda} \right)\hat{\sigma}_{3}
\hat{\psi}_{k\lambda\sigma}(t)
+\sum_{\sigma}\hat{\Phi}_{\sigma}^{\dagger}(t)\left(i\partial_{t} - E_{d\sigma} \right)\hat{\sigma}_{3}
 \hat{\Phi}_{\sigma}(t)\right.\nonumber\\
&&\left.
+\sum_{k,\lambda,\sigma}\left[V_{k\lambda}\hat{\Phi}_{\sigma}^{\dagger}(t)\hat{\sigma}_{3}\hat{\psi}_{k\lambda\sigma}(t) + h.c. \right] \right\}
+ i\,S^{int}[\hat{\Phi}^{\dagger},\hat{\Phi};U].
\label{eqAC2}
\end{eqnarray}
Here, the fields are two-component spinors in the index that distinguishes between the forward and backward branch of the Keldysh contour and $\sigma_3$ is the 3rd Pauli matrix in this space.
The index $\lambda=L,\,R$ represents the left and right electrodes, $U$ is the Coulomb interaction, and $E_{d\sigma} = \epsilon_{d\sigma} + U/2$ is the shift of the local resonance level
with respect to the particle-hole symmetric condition $\epsilon_{d} = -U/2$. In the presence of a local magnetic field $B$, the local level has to be shifted by the Zeeman term,
{\it i.e.}, $\epsilon_{d\sigma} = \epsilon_{d}-\sigma g \mu_{B}B/2$. 
The action defined by Eq.(\ref{eqAC2}) is Gaussian in the Grassman fields $\hat{\psi}_{k\lambda\sigma}$ representing the lead electrons by virtue of their non-interacting nature.
As a result, the $\hat{\psi}_{k\lambda\sigma}$ fields can be integrated out. This procedure yields an effective action for the localized electron states described by the Grassman fields $\hat{\Phi}_{\sigma\omega}$, which represent the $d_{\sigma}$ of Eq.~(\ref{Andersonmodel}).
In frequency-space this effective action is given by~\cite{Munoz.13}
\begin{eqnarray}
i S[\hat{\Phi}^{\dagger},\hat{\Phi};U,E_{d\sigma},\Delta] &=& i\int_{-\infty}^{+\infty}\frac{d\omega}{2\pi}
\sum_{\sigma}\hat{\Phi}_{\sigma\omega}^{\dagger}\left(\omega + i\Delta \right)\hat{\sigma}_{3}
\hat{\Phi}_{\sigma\omega} +\, i S^{int}[\hat{\Phi}^{\dagger},\hat{\Phi};U]\nonumber\\
 &&- i\int_{-\infty}^{+\infty}\frac{d\omega}{2\pi}\sum_{\sigma}
\hat{\Phi}_{\sigma\omega}^{\dagger}E_{d\sigma}\hat{\sigma}_{3}\hat{\Phi}_{\sigma\omega}.
\label{eqAC3}
\end{eqnarray}
Here, we have defined $\Delta = \Delta_{L} + \Delta_{R}$, and
\begin{eqnarray}
i \Delta_{\lambda} = -\sum_{k,\lambda}\frac{|V_{k\lambda}|^{2}}{\omega - \epsilon_{k\lambda} + i\eta^{+}}.
\label{eqAC4}
\end{eqnarray}
In analogy to the equilibrium version of the renormalized theory (RPT) of Hewson~\cite{Hewson.01,Hewson.06,Hewson.11}, we can express the action Eq.(\ref{eqAC3})
in terms of a set of "renormalized" parameters $\tilde{U}$, $\tilde{\epsilon}_{d\sigma}$ and $\tilde{\Delta}$. The fields are rescaled accordingly via the wave-function renormalization factor $z$ given by
$z^{-1} = 1 - \left.\partial\Sigma^{R}_{\sigma\omega}/\partial\omega\right|_{\omega=0}$. 

Here, we seek to construct a renormalized perturbation theory to treat deviations from particle-hole symmetry, represented
by the paramater $\tilde{\epsilon}_{d\sigma} = z E_{d\sigma}$ which is assumed to be small 
compared to the energy scale determined by
the renormalized quasi-particle spectral width $\tilde{\Delta} = z \Delta$. 
Thus,
\begin{eqnarray}
S[\hat{\Phi}^{\dagger},\hat{\Phi};U,E_{d\sigma},\Delta] &=& S[\tilde{\Phi}^{\dagger},\tilde{\Phi};\tilde{U},\tilde{\epsilon}_{d\sigma},\tilde{\Delta}] 
+ \delta S[\tilde{\Phi}^{\dagger},\tilde{\Phi};\lambda_{1},\lambda_{2},\lambda_{3}].
\label{eqAC5}
\end{eqnarray}
In order to preserve the original action and thus avoid over counting, one must include  counterterms proportional to the parameters $\lambda_{1}$, $\lambda_{2}$ and $\lambda_{3}$.
These counterterms are defined as $\lambda_{1} = -z\Sigma_{\sigma}^{R}(0,0)$,
$\lambda_{2} = z-1$ and $\lambda_{3} = z^{2}\left( U - \Gamma_{\uparrow\downarrow}(0,0)  \right)$,
respectively. Their values are determined in order to satisfy the RG conditions imposed on the retarded Green's function
\begin{eqnarray}
\left.\tilde{\Sigma}_{\sigma\omega}^{R}\right|_{\omega=0,T=0,V=0} &=& 0,\nonumber\\
\left.\frac{\partial}{\partial\omega}\tilde{\Sigma}_{\sigma\omega}^{R}\right|_{\omega=0,T=0,V=0} &=& 0,\nonumber\\
\tilde{\Gamma}_{\uparrow\downarrow}(0,0)=z^{2}\Gamma_{\uparrow\downarrow}(0,0) &=& \tilde{U},
\label{eqAC6}
\end{eqnarray}
at every order in $\tilde{U}$.
For the particle-hole symmetric case, the retarded
self-energy in equilibrium has been obtained up to second order in $\tilde{u} = \tilde{U}/(\pi\tilde{\Delta})$~\cite{Hewson.01,Hewson.06,Hewson.11}.  This equilibrium expression can be extended to non-thermal steady states, {\it e.g.} the one that ensues when the system is experiencing a voltage drop across the interaction region that is constant in time. This extension is based on Ward identities \cite{Oguri.01,Oguri.05,Munoz.13,suppl}. As a result, one obtains an expression for the selfenergy of the local propagator near the strong-coupling fixed point and at particle-hole symmetry and zero magnetic field
\begin{eqnarray}
\tilde{\Sigma}_{\sigma\omega}^{R} &=& -i\frac{\tilde{\Delta}\tilde{u}^{2}}{2}\left[\left(\frac{\omega}{\tilde{\Delta}} \right)^{2} + \left(\frac{\pi T}{\tilde{\Delta}} \right)^{2} + \zeta\left(\frac{V}{\tilde{\Delta}} \right)^{2}-\frac{\zeta}{3}\left(\frac{\pi T V}{\tilde{\Delta}} \right)^{2} \right] + O(T^4,V^4,\tilde{u}^{3}).
\label{eqSE1}
\end{eqnarray}

At zero magnetic field, it has been shown by Hewson \cite{Hewson.01,Hewson.11} that the counterterm $\lambda_{1}$
cancels the Hartree contribution up to order $\tilde{u}^{2}$. To the same order,
there are no contributions from the counterterms $\lambda_{2}$ and $\lambda_{3}$ \cite{Hewson.01,Hewson.11}. 
This renormalized action, exact up to terms of order $\tilde{u}^{2}$
constitutes the reference system around which  our perturbative scheme of dual fermions is defined.
In the next section, the bare dual fermion Green's function is introduced.

\section{The dual Fermion Green's function}

The bare dual fermion bare Green's function is given by the expression \cite{Munoz.13,suppl}
\begin{eqnarray}
\mathbf{G}_{\sigma\omega}^{f(0)} &=& -\mathbf{g}_{\sigma\omega}\left[\mathbf{g}_{\sigma\omega} - \mathbf{D}_{\sigma\omega}^{-1} \right]^{-1}\mathbf{g}_{\sigma\omega} = \sum_{n=1}^{\infty}\mathbf{g}_{\sigma\omega}\left[\mathbf{D}_{\sigma\omega}\mathbf{g}_{\sigma\omega} \right]^{n}.
\label{eqDF1}
\end{eqnarray}
Here, we have defined $\mathbf{D}_{\sigma\omega} = \tilde{\epsilon}_{d\sigma}\hat{\sigma}_{3}$. For convenience we switch from the dynamical index representation $+-$ to the trigonal representation~\cite{Kamenev2} in what follows. Both representations are linked by a similarity transformation
\begin{eqnarray}
\hat{\mathbf{G}}^{f}_{\sigma\omega}= \hat{L}\hat{\sigma}_{3}\mathbf{G}^{f}_{\sigma\omega}\hat{L}^{\dagger} = \left(\begin{array}{cc}G^{f,R}_{\sigma\omega} & G^{f,K}_{\sigma\omega}\\0 & G^{f,A}_{\sigma\omega} \end{array}\right),
\label{eqDF2}
\end{eqnarray}
where we defined $\hat{L} = \left(1 - i\hat{\sigma}_{2} \right)/\sqrt{2}$ and $\hat{\sigma}_{2}$ is the 2nd Pauli matrix.
Under this transformation the matrix $\mathbf{D}_{\sigma\omega}\rightarrow \hat{L}\hat{\sigma}_{3}\mathbf{D}_{\sigma\omega}\hat{L}^{\dagger} = \tilde{\epsilon}_{d\sigma} \mathbf{1} $ becomes proportional to the identity. Moreover, the retarded component of the dual fermion Green's function becomes
\begin{eqnarray}
G^{f(0),R}_{\sigma\omega} = \sum_{n=1}^{\infty}\left(\tilde{\epsilon}_{d\sigma}\right)^{n}
\left(g_{\sigma\omega}^{R}\right)^{n+1}=\frac{ \tilde{\epsilon}_{d\sigma}\,\left(g_{\sigma\omega}^{R}\right)^{2}}{1 - \tilde{\epsilon}_{d\sigma}g_{\sigma\omega}^{R}}.
\label{eqDF3}
\end{eqnarray}
The advanced Green's function then follows from  the usual relation $G^{f(0),A}_{\sigma\omega}=\left(G^{f(0),R}_{\sigma\omega}\right)^{*}$.
In the  steady-state  and up to $O(\tilde{u}^{4})$, the Keldysh component of the dual fermion Green's function is given by \cite{Munoz.13,suppl}
\begin{eqnarray}
G^{f(0),K}_{\sigma\omega} = \left(1 - 2 \tilde{f}(\omega,T,V) \right)\left(G^{f(0),A}_{\sigma\omega} - G^{f(0),R}_{\sigma\omega} \right).
\label{eqDF4}
\end{eqnarray}

Here, up to $O(\tilde{u}^{4})$, the non-equilibrium distribution function is given  by the expression \cite{Munoz.13,suppl}
\begin{eqnarray}
\tilde{f}(\omega,T,V) = \frac{\tilde{\Delta}_{L}f_{0}(\omega+\alpha_{L}V,T) + \tilde{\Delta}_{R}f_{0}(\omega-\alpha_{R}V,T)}{\tilde{\Delta}_{L} + \tilde{\Delta}_{R}},
\label{eqDF5}
\end{eqnarray}
where $f_{0}(\omega,T)$ denotes the Fermi distribution and $\alpha_{L}+\alpha_{R}=1$. 

Note, that by substituting the definition of the retarded renormalized
Green's function of the reference system 
\begin{eqnarray}
g_{\sigma\omega}^{R}=\left(\omega + i\tilde{\Delta}- \tilde{\Sigma}_{\sigma\omega}^{R}\right)^{-1}
\label{eqDF6}
\end{eqnarray}
into Eq.(\ref{eqDF3}), one concludes that as $\tilde{\epsilon}_{d\sigma}\rightarrow 0$ then $G^{f(0),R}_{\sigma\omega}\rightarrow 0$, and hence the dual fermion contribution
to the perturbed self-energy vanishes in the p-h symmetric limit, as expected.

On the other hand, in the limit $\tilde{\epsilon}_{d\sigma} \rightarrow \infty$,
we have the non-trivial result~\cite{Rubtsov.08,Munoz.13}
\begin{eqnarray}
\lim_{\tilde{\epsilon}_{d\sigma\rightarrow}\infty}G^{f(0),R}_{\sigma\omega} = -g_{\sigma\omega}^{R}.
\label{eqDF7}
\end{eqnarray}

The relation between the dual fermion Green's function and the local Green's function associated with the $d^{\dagger}_{\sigma},d^{}_{\sigma}$ operators of Eq.~(\ref{Andersonmodel}) in Keldysh space is
\begin{eqnarray}
\mathbf{G}_{\sigma\omega} = -\tilde{\epsilon}_{d\sigma}^{-1}\hat{\sigma}_{3} + \left(\mathbf{g}_{\sigma\omega}\hat{\sigma}_{3}\tilde{\epsilon}_{d\sigma} \right)^{-1}\mathbf{G}_{\sigma\omega}^{f}\left(\hat{\sigma}_{3}\mathbf{g}_{\sigma\omega}\tilde{\epsilon}_{d\sigma} \right)^{-1}.
\label{eqDF7b}
\end{eqnarray}
Here, the dressed dual fermion Green's function $\mathbf{G}_{\sigma\omega}^{f}$ is a solution of the Dyson equation
\begin{eqnarray}
\mathbf{G}_{\sigma\omega}^{f} = \mathbf{G}_{\sigma\omega}^{f(0)} + \mathbf{G}_{\sigma\omega}^{f(0)}\mathbf{\Sigma}_{\sigma\omega}^{f}\mathbf{G}_{\sigma\omega}^{f},
\label{eqDF8}
\end{eqnarray}
where $\mathbf{\Sigma}_{\sigma\omega}^{f}$ is the dual fermion self-energy which will be discussed in the next section.

It is convenient to express Eqs.~(\ref{eqDF7b}) in the trigonal representation, by applying the transformation defined in Eq.(\ref{eqDF2}), which leads to
\begin{eqnarray}
\hat{\mathbf{G}}_{\sigma\omega} = \hat{L}\hat{\sigma}_{3}\mathbf{G}_{\sigma\omega}\hat{L}^{\dagger} = -\tilde{\epsilon}_{d\sigma}^{-1}\mathbf{I} + \tilde{\epsilon}_{d\sigma}^{-2}\hat{\mathbf{g}}_{\sigma\omega}^{-1}\hat{\mathbf{G}}_{\sigma\omega}^{f}\hat{\mathbf{g}}_{\sigma\omega}^{-1}.
\label{eqDF9}
\end{eqnarray}
Similarly, the Dyson equation becomes in the trigonal representation
\begin{eqnarray}
\hat{\mathbf{G}}_{\sigma\omega}^{f} = \hat{\mathbf{G}}_{\sigma\omega}^{f(0)} + \hat{\mathbf{G}}_{\sigma\omega}^{f(0)}\hat{\mathbf{\Sigma}}_{\sigma\omega}^{f}\hat{\mathbf{G}}_{\sigma\omega}^{f},
\label{eqDF10}
\end{eqnarray}
where the dual-fermion self-energy adopts the matrix form
\begin{eqnarray}
\hat{\mathbf{\Sigma}}_{\sigma\omega}^{f} = \left(\begin{array}{cc}\Sigma^{f,R}_{\sigma\omega}&\Sigma^{f,K}_{\sigma\omega}\\0 &\Sigma^{f,A}_{\sigma\omega}\end{array}\right).
\label{eqDF11}
\end{eqnarray}

In particular, solving for the retarded component of the non-equilibrium local Green's function from Eq.(\ref{eqDF7b}), we have
\begin{eqnarray}
G_{\sigma\omega}^{R} = -\tilde{\epsilon}_{d\sigma}^{-1} + \tilde{\epsilon}_{d\sigma}^{-2}\left( g_{\sigma\omega}^{R} \right)^{-1} G_{\sigma\omega}^{f,R} \left( g_{\sigma\omega}^{R} \right)^{-1}.
\label{eqDF12}
\end{eqnarray}

This exact expression can be combined with the solution of the Dyson equation for the retarded component of the dual-fermion Green's function obtained from Eq.(\ref{eqDF8}),
\begin{eqnarray}
G_{\sigma\omega}^{f,R} = \frac{G_{\sigma\omega}^{f(0),R}}{1 - G_{\sigma\omega}^{f(0)}\Sigma_{\sigma\omega}^{f,R}},
\label{eqDF13}
\end{eqnarray}
to obtain an explicit expression for the retarded component of the local non-equilibrium Green's function
\begin{eqnarray}
G_{\sigma\omega}^{R} = \left(\omega - \tilde{\epsilon}_{d\sigma} + i\tilde{\Delta} - \tilde{\Sigma}_{E_{d},\sigma\omega}^{R} \right)^{-1}.
\label{eqDF14}
\end{eqnarray}
Here, we have defined the retarded component of the non-equilibrium self-energy by
\begin{eqnarray}
\tilde{\Sigma}_{E_{d},\sigma\omega}^{R} = \tilde{\Sigma}_{\sigma\omega}^{R} + \frac{\Sigma_{\sigma\omega}^{f,R}}{1 + g_{\sigma\omega}^{R}\Sigma_{\sigma\omega}^{f,R}}.
\label{eqDF15}
\end{eqnarray}

\section{The dual fermion self-energy}

In order to obtain  an approximation for the dual fermion self-energy, we consider the sum 
of ladder diagrams, with the effective quasi-particle interaction defined by the renormalized
vertex $\tilde{U}$ of Eq.~(\ref{eqAC6}). 
In the trigonal representation, the non-equilibrium dual-fermion
self-energy matrix, expressed in terms of the dual fermion vertex $\hat{\mathbf{\Gamma}}^{f}$, is
\begin{eqnarray}
\hat{\mathbf{\Sigma}}_{\sigma\omega}^{f} = \frac{1}{2}\int_{-\infty}^{+\infty}\frac{d\omega'}{2 i \pi}\left\{\Gamma^{f,K}_{\sigma,-\sigma}\left(\hat{\gamma}^{1}\hat{\mathbf{G}}^{f(0)}_{-\sigma\omega'}\hat{\gamma}^{1}\right) +
\Gamma^{f,A}_{\sigma,-\sigma}\left(\hat{\gamma}^{1}\hat{\mathbf{G}}^{f(0)}_{-\sigma\omega'}\hat{\gamma}^{2}\right) +
\Gamma^{f,R}_{\sigma,-\sigma}\left(\hat{\gamma}^{2}\hat{\mathbf{G}}^{f(0)}_{-\sigma\omega'}\hat{\gamma}^{1}\right)
\right\},
\label{eqDFSE1}
\end{eqnarray}
where  $\hat{\gamma}^{1} = \mathbf{1}$ and $\hat{\gamma}^{2} = \hat{\sigma}_{1}$.

Here, the dual-fermion vertex components are defined by the matrix
\begin{eqnarray}
\hat{\mathbf{\Gamma}}^{f}_{\sigma,-\sigma} = \left[\mathbf{1} -  \tilde{U}\hat{\sigma}_{1}\hat{\mathbf{\Pi}}^{f(0)}_{\sigma,-\sigma}(\omega)\right]^{-1}\hat{\sigma}_{1}\tilde{U} = \left(\begin{array}{cc}\Gamma^{f,K}_{\sigma,-\sigma} & \Gamma^{f,R}_{\sigma,-\sigma}\\
\Gamma^{f,A}_{\sigma,-\sigma} & 0 \end{array} \right), 
\label{eqDFSE2}
\end{eqnarray}
where the polarisation matrix is defined by
\begin{eqnarray}
\hat{\mathbf{\Pi}}^{f(0)}_{\sigma,-\sigma}(\omega) = \left(\begin{array}{cc} 0 & \Pi^{A}_{\sigma,-\sigma}(\omega) \\ \Pi^{R}_{\sigma,-\sigma}(\omega)& \Pi^{K}_{\sigma,-\sigma}(\omega) \end{array}\right). 
\label{eqDFSE3}
\end{eqnarray}

The different components
of the polarization insertion in the trigonal representation are given by the expressions \cite{Kamenev2}
\begin{eqnarray}
\Pi^{f,K}_{\sigma,-\sigma}(\omega) &=& -\frac{1}{2}\int_{-\infty}^{+\infty}
\frac{d\omega'}{2 i\pi} \left\{
G_{\sigma,\omega + \omega'}^{f(0),K}G_{-\sigma\omega'}^{f(0),K}
+ G_{\sigma,\omega + \omega'}^{f(0),R}G_{-\sigma\omega'}^{f(0),A}
+ G_{\sigma,\omega + \omega'}^{f(0),A}G_{-\sigma\omega'}^{f(0),R}
\right\},\\
\Pi^{f,R}_{\sigma,-\sigma}(\omega) &=& -\frac{1}{2}\int_{-\infty}^{+\infty}
\frac{d\omega'}{2 i\pi} \left\{ G_{\sigma,\omega + \omega'}^{f(0),R}G_{-\sigma\omega'}^{f(0),K} + G_{\sigma,\omega + \omega'}^{f(0),K}
G_{-\sigma\omega'}^{f(0),A} \right\},\\
\Pi^{f,A}_{\sigma,-\sigma}(\omega) &=& \left[\Pi^{f,R}_{\sigma,-\sigma}(\omega)\right]^{*}.
\label{eqDFP15}
\end{eqnarray}
In particular, at $\omega =0$, $T =0$, $V = 0$, and to zero order in $\tilde{U}$, one has
\begin{eqnarray}
\Pi^{f(0),R}_{\sigma,-\sigma}(0)= \Pi^{f(0),A}_{\sigma,-\sigma}(0) = \frac{1}{\pi\tilde{\Delta}} + \frac{1}{\pi\tilde{\Delta}}\frac{\tilde{\epsilon}_{d,-\sigma}^{2}{\rm{tan}}^{-1}(\tilde{\epsilon}_{d\sigma}) - \tilde{\epsilon}_{d\sigma}^{2}{\rm{tan}}^{-1}(\tilde{\epsilon}_{d,-\sigma})}{\tilde{\epsilon}_{d\sigma}\tilde{\epsilon}_{d,-\sigma}\left(\tilde{\epsilon}_{d\sigma}
-\tilde{\epsilon}_{d,-\sigma} \right)}.
\label{eqDFP16}
\end{eqnarray}

The dual fermion vertex components are thus obtained as
\begin{eqnarray}
\Gamma^{f,R}_{\sigma,-\sigma} &=& \frac{\tilde{U}_{s}}{1 - \tilde{U}_{s}\Pi^{f,R}_{\sigma-\sigma}(\omega)},\nonumber\\
\Gamma^{f,A}_{\sigma,-\sigma} &=& \frac{\tilde{U}_{s}}{1 - \tilde{U}_{s}\Pi^{f,A}_{\sigma,-\sigma}(\omega)},\nonumber\\
\Gamma^{f,K}_{\sigma,-\sigma} &=& -\frac{\tilde{U}_{s}^{2}\Pi^{f,K}_{\sigma,-\sigma}(\omega)}{\left(1 -
\tilde{U}_{s}\Pi_{\sigma,-\sigma}^{f,R}(\omega) \right)\left(1 - \tilde{U}_{s}\Pi_{\sigma,-\sigma}^{f,A}(\omega) \right)}.
\label{eqDFV1}
\end{eqnarray}

Note that the Keldysh component of the dual fermion vertex is of higher order
in the effective quasiparticle interaction $\tilde{U}_{s}$ than
the retarded and advanced components.

In order to obtain analytical approximations to the transport coefficients, defined in Eq.~(\ref{eqIntro1}), which are exact up to $O(\tilde{u}^{3})$, we define an effective dual-fermion interaction from the expression for the dual-fermion
vertex at $\omega=0$, $V=0$, and $T=0$ by
\begin{eqnarray}
\tilde{U}^{f}_{\sigma,-\sigma} =\left. \Gamma_{\sigma,-\sigma}^{f,A}\right|_{\omega=0,V=0,T=0} = \left.\Gamma_{\sigma,-\sigma}^{f,R}\right|_{\omega=0,V=0,T=0} = \frac{\tilde{U}_{s}}{1 - \tilde{U}_{s}\Pi_{\sigma,-\sigma}^{f,R}(0)}.
\label{eqDFU}
\end{eqnarray}

Up to  order $O(\tilde{u}^{3})$ 
we have that the self-energy of the non-symmetric system defined by Eq.(\ref{eqDF15})
reduces to the expression
\begin{eqnarray}
\tilde{\Sigma}_{E_{d},\sigma\omega}^{R} &=& \tilde{\Sigma}_{\sigma\omega}^{R} + \Sigma_{\sigma\omega}^{f,R} + O(\tilde{u}^{3}).
\label{eqSE2}
\end{eqnarray}

\section{Transport coefficients}

With the results of the previous sections, we are in a position to calculate the experimentally accessible transport coefficients $\alpha$ and $\gamma$ defined in Eq.~(\ref{eqIntro1}).
From the Meir-Wingreen expression \cite{Meir.92} for the charge current  through the quantum dot,
\begin{eqnarray}
I = \frac{e}{2\hbar}\sum_{\sigma}\int  d\omega  \frac{4\Delta_{L}\Delta_{R}}{\Delta_{L} + \Delta_{R}}   A_{\sigma}(\omega,T,V)\left[ f_{L}(\omega) - f_{R}(\omega) \right] 
\label{eqI1}
\end{eqnarray}
where $f_{L,R} = f(\omega - \mu_{L,R})$ is the Fermi function for the left/right lead, respectively,
and $A_{\sigma}(\omega,T,V) = -\pi^{-1}\Im\, G_{\sigma\omega}^{R}$ is the local spectral function. 
Expanding the differential conductance $dI/dV \equiv G(T,V,B)$ up to second order in $V$, $T$, and $B$, one obtains the expression \cite{Munoz.13}
\begin{equation}
\frac{G(T,0,B) - G(T,V,B)}{G(0,0,0)} = c_{V}\left(\frac{eV}{\tilde{\Delta}} \right)^{2} - c_{TV}\left(\frac{eV}{\tilde{\Delta}}  \right)^{2}\left( \frac{k_{B}T}{\tilde{\Delta}}  \right)^{2} 
-c_{V E_{d}}\left( \frac{eV}{\tilde{\Delta}} \right) + c_{TVE_{d}}\left( \frac{eV}{\tilde{\Delta}} \right)^{2}.
\label{eqI2}
\end{equation}

\begin{figure}[h]
\begin{center}
\includegraphics[width=0.7\linewidth]{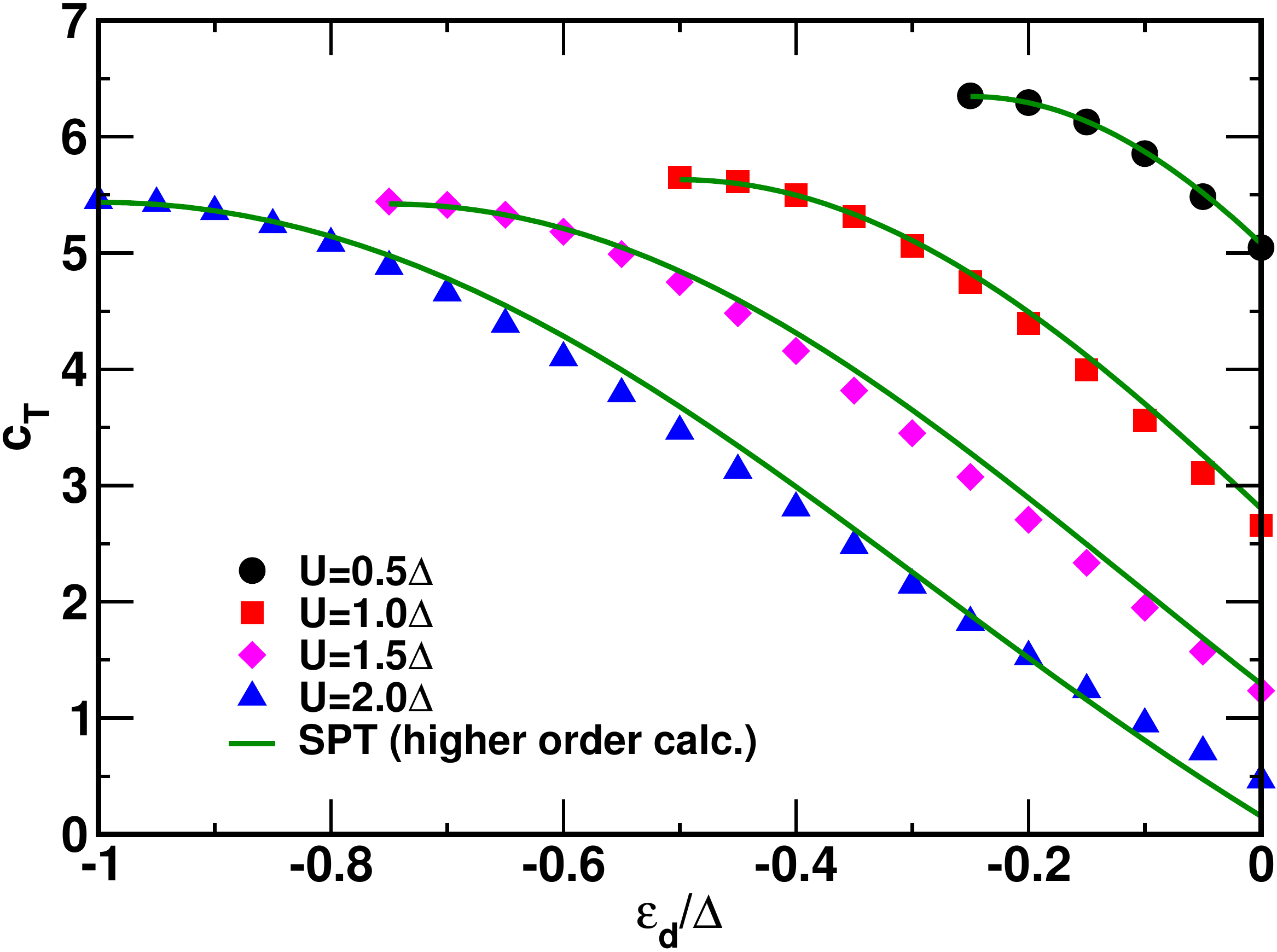}
\caption{\label{figure1}The transport  coefficient $c_T$ versus the asymmetry parameter $\epsilon_d/\Delta$: Continuous lines are the renormalized SPT results including  higher-order terms. Dashed lines are the renormalized SPT results at the lowest order. Symbols are the NRG results for the same set of parameters. The particle-hole symmetric reference system is located at $\epsilon_d=-U/2$.}
\end{center}
\end{figure}

\begin{figure}[h]
\begin{center}
\includegraphics[width=0.7\linewidth]{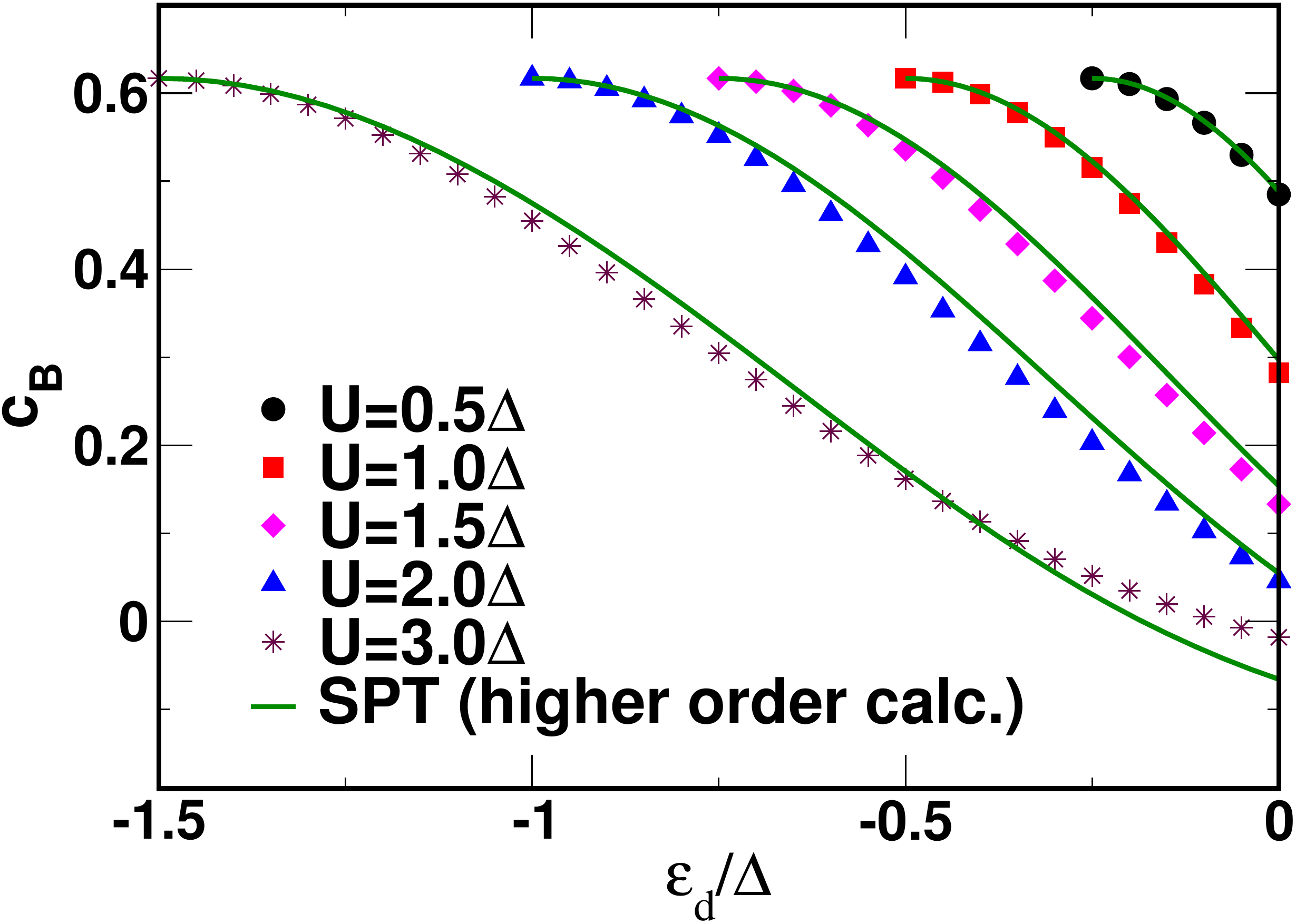}
\caption{\label{figure2}The transport coefficient $c_{B}$ as a function of the asymmetry parameter $\epsilon_d/\Delta$: Continuous lines are the renormalized SPT results including  higher-order terms.
Dashed lines are the renormalized SPT results at the lowest order. Symbols are the NRG results for the same set of parameters.}
\end{center}
\end{figure}

In particular, for the linear-response conductance ($V\rightarrow 0$), we have the relations
\begin{eqnarray}
\frac{G(T,V=0,B=0)}{G(0,0,0)} &=& 1 - c_{T}\left( \frac{k_{B}T}{\tilde{\Delta}}  \right)^{2},\nonumber\\
\frac{G(T=0,V=0,B)}{G(0,0,0)} &=& 1 - c_{B}\left( \frac{g\mu_{B}B}{\tilde{\Delta}}  \right)^{2}.
\label{eqCTB1}
\end{eqnarray}

In Fig.~\ref{figure1}, we show the renormalized SPT result for $c_{T}$, defined in Eq.(\ref{eqCTB1}),  obtained by including the
higher-order dual-fermion contributions as described in section 4. 
The linear-response conductance can also be calculated via the NRG, which yields essentially exact results for this quantity~\cite{Costi.94c,Merker.13}.
 Clearly, the renormalized
SPT results for $c_T$ are in excellent  agreement with NRG calculations, even at relatively large values of the asymmetry parameter $\epsilon_{d}/\Delta$ 
and Coulomb interaction $U$. We also display the results (dashed lines in Fig.~\ref{figure1}) obtained from the SPT calculation at the lowest order \cite{Merker.13}, where the improvement upon inclusion
of higher order contributions is evident.
A similar level of agreement with the NRG is obtained for the transport coefficient $c_B$, as shown in Fig.~\ref{figure2}. 

\section{Conclusion}

In summary, we have reviewed the renormalized SPT approach to non-thermal steady states in the Anderson model. By construction, this approach captures the strong-coupling nature of the fixed point.
We present an extension  to higher orders and compared these higher-order results to NRG calculations. 
As a result, the renormalized SPT predictions are in excellent  agreement with the (essentially) exact NRG results. The renormalized SPT is a versatile approach for the out-of-equilibrium properties at sufficiently low energies and temperatures that can also be applied to more complicated models or more general steady states, that ensue by applying finite voltage and temperature differences across the system ~\cite{Kirchner.12NL}. We also elucidated the role of Ward identities in calculating the out-of-equilibrium properties and clarified several controversial statements that appeared in the literature.

\section*{Acknowledgments}

E.\,Mu\~{n}oz  acknowledges financial support by Fondecyt (Chile) under contract No.1141146.
S.~Kirchner acknowledges partial support by the National Science Foundation of China, grant No.11474250, and the National Key R\&D Program of the MOST of China (No.2016YFA0300202).

\section*{Appendix A: Violation of Oguri's Ward identity in certain  RPT schemes}

\begin{figure}[h!]
\begin{center}
\includegraphics[width=5.0cm]{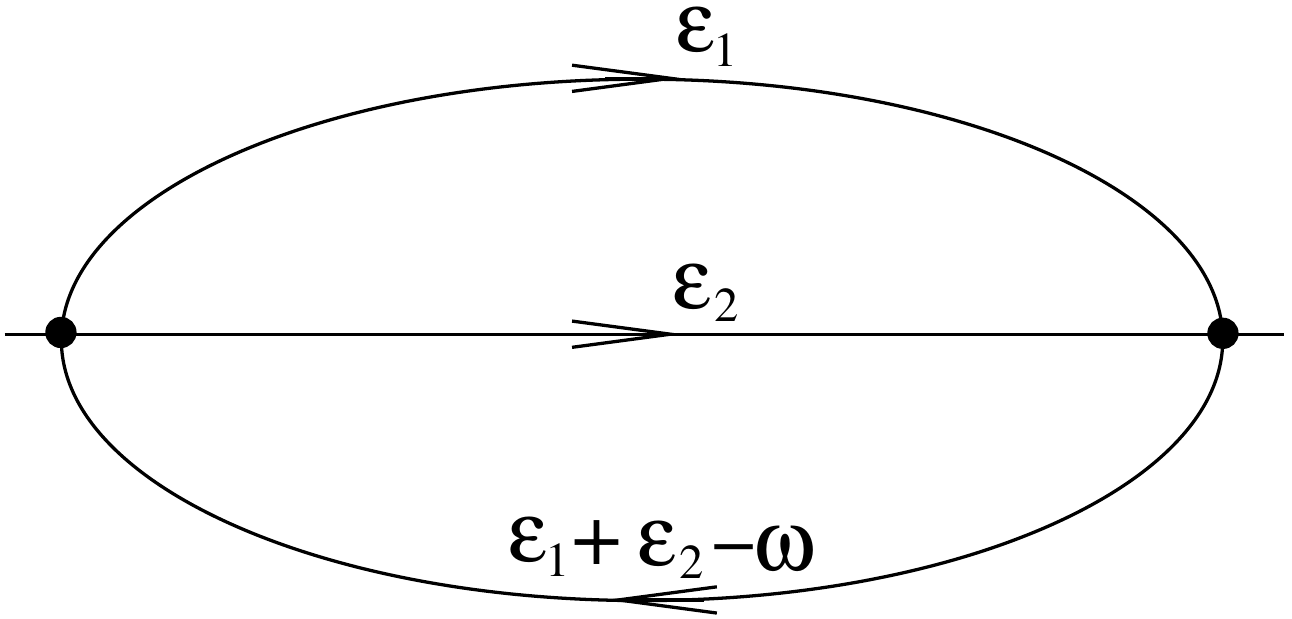}
\end{center}
\caption{The sunset diagram appears in second order in the renormalized coupling constant and is part of the 'reference system' of the renormalized SPT approach.
}
\label{fig:A1}
\end{figure}

In this appendix we briefly clarify a few controversial statements that appeared in the literature on the validity and the range of applicability of Oguri's Ward identity. 
Particle number conservation and the conservation of spin in a magnetic field lead to relations among derivatives of the proper local selfenergy $\Sigma(\omega)$ of the Anderson Hamiltonian, Eq.~(\ref{Andersonmodel}), with respect to $E_{d\sigma}$, magnetic field $B$, and $\omega$ through Ward identities. These identities were first obtained by Yamada and Yosida~\cite{Yamada.75}.
The extensions to finite voltages is due to Oguri~\cite{Oguri.01} who showed that
\begin{eqnarray}
\label{OguriWard}
\left.\frac{\partial}{\partial V}\boldsymbol{\tilde{\Sigma}}^{}(\omega,T,V)\right|_{V=0} = - \gamma \left( \frac{\partial}{\partial\omega} + \frac{\partial}{\partial \tilde{\epsilon}_{d}}  \right) \boldsymbol{\tilde{\Sigma}}^{}(\omega,T,V=0),
\end{eqnarray}
here given in terms of renormalized quantities and where $\boldsymbol{\tilde{\Sigma}}^{}$ is the selfenergy matrix on the Keldysh contour as in Eq.(\ref{eqDF11}). The parameter $\gamma = \left(\Delta_{L}\alpha_{L} - \Delta_{R}\alpha_{R}\right)/\left(\Delta_{L} + \Delta_{R}\right)$.
To demonstrate that the relation (\ref{OguriWard}) is valid at each order of perturbation theory, we here 
explicitly demonstrate that the sunset diagram, shown in Fig.~\ref{fig:A1}, fulfills Oguri's Ward identity. A general proof can be obtained following similar arguments~\cite{Oguri.05}.

Specifically, the lesser component of $\boldsymbol{\tilde{\Sigma}}$ is considered which, 
on the Keldysh contour,  is simply given by
\begin{equation}	
\tilde{\Sigma}^<(\tau)=\tilde{U}^2\tilde{G}^{<}(\tau)\tilde{G}^{<}(\tau)\tilde{G}^{>}(-\tau)
\label{eqAA1}
\end{equation}
or after Fourier transformation to frequency variables
\begin{eqnarray}
\tilde{\Sigma}^{<}(\omega,T,V)&=&-2\pi i \tilde{U}^2\int d\epsilon_1\int d\epsilon_2 \tilde{f}(\epsilon_1)\tilde{f}(\epsilon_2) \tilde{f}(\omega-\epsilon_1-\epsilon_2)\tilde{A}(\epsilon_1)\tilde{A}(\epsilon_2)\tilde{A}(\epsilon_1+\epsilon_2-\omega),\nonumber
\label{eqAA2}
\end{eqnarray}	
where  
\begin{equation}
\label{specdensity}
\tilde{A}(x)=\frac{1}{\pi}\frac{\tilde{\Delta}}{(x-\tilde{\epsilon}_d)^2+\tilde{\Delta}^2}
\end{equation}
is the bare local renormalized spectral density which obeys the identity
\begin{equation}
\left (\frac{\partial}{\partial x}+\frac{\partial}{\partial \tilde{\epsilon}_d}\right ) \tilde{A}(x)=0.
\label{eqAA3}
\end{equation}
Note that here, $\tilde{\epsilon}_{d} = z E_{d}$  is the renormalized, voltage and temperature independent local resonance level.
The effective distribution function $\tilde{f}(\omega,V)$ is given by the expression
\begin{eqnarray}
\tilde{f}(\omega,V) = \sum_{\lambda=L,R}\frac{\Delta_{\lambda}}{\Delta_{L}+\Delta_{R}}f_{0}(\omega-\alpha_{\lambda}V),
\label{eqAA4}
\end{eqnarray}
where the chemical potential at lead $\lambda=L,R$ is $\mu_{L}=\alpha_{L}V$, $\mu_{R} = -\alpha_{R}V$, with $\alpha_{L}+\alpha_{R}=1$.
Equation (\ref{eqAA1}) implies
\begin{eqnarray}
\lim_{V\rightarrow 0}\tilde{f}(\omega,V) &=& f_{0}(\omega),\nonumber\\
\left.\frac{\partial \tilde{f}(\epsilon,V)}{\partial V}\right|_{V=0} &=& \gamma \left(-\frac{\partial f_{0}(\epsilon)}{\partial\epsilon} \right),
\label{eqAA5}
\end{eqnarray}
where $f_0$ is the usual Fermi function.

Consider the voltage derivative of the self-energy expression in Eq.~(\ref{eqAA1}), which 
can be written 
\begin{eqnarray}
\left.\frac{\partial}{\partial V}\tilde{\Sigma}^{<}(\omega,T,V)\right|_{V=0} = -2\pi i \tilde{U}^{2} \gamma\left( I_{1} + I_{2} + I_{3}  \right),
\label{eqAA6}
\end{eqnarray}
with the help of Eq.~(\ref{eqAA5}) and where we have defined
\begin{eqnarray}
I_{1} &=& \int d\epsilon_{1} \int d\epsilon_{2} \tilde{A}(\epsilon_{2})\tilde{A}(\epsilon_{1} + \epsilon_{2} - \omega)f_{0}(\epsilon_{2})f_{0}(\omega - \epsilon_{1} - \epsilon_{2})\left\{
\tilde{A}(\epsilon_{1})\left( -\frac{\partial f_{0}(\epsilon_{1})}{\partial\epsilon_{1}}  \right)
\right\}\nonumber\\
I_{2} &=& \int d\epsilon_{1} \int d\epsilon_{2} \tilde{A}(\epsilon_{1})\tilde{A}(\epsilon_{1} + \epsilon_{2} - \omega)f_{0}(\epsilon_{1})f_{0}(\omega - \epsilon_{1} - \epsilon_{2})\left\{
\tilde{A}(\epsilon_{2})\left( -\frac{\partial f_{0}(\epsilon_{2})}{\partial\epsilon_{2}}  \right)\right\}\nonumber\\
I_{3} &=& \int d\epsilon_{1} \int d\epsilon_{2} \tilde{A}(\epsilon_{1})\tilde{A}(\epsilon_{2})f_{0}(\epsilon_{1})f_{0}(\epsilon_{2})\left\{
\tilde{A}(\epsilon_{1}+\epsilon_{2}-\omega)\left( -\frac{\partial}{\partial\epsilon_{1}} f_{0}(\omega-\epsilon_{1}-\epsilon_{2})  \right)
\right\}.
\label{eqAA7}
\end{eqnarray}
We now focus on the term $I_{1}$ of Eq.\~(\ref{eqAA6}) and
apply the identity
\begin{eqnarray}
\tilde{A}(\epsilon_{1})\left(-\frac{\partial f_{0}(\epsilon_{1})}{\partial\epsilon_{1}}  \right)
= -\tilde{A}(\epsilon_{1})\left( \frac{\partial}{\partial\epsilon_{1}} + \frac{\partial}{\partial \tilde{\epsilon}_{d}}  \right)f_{0}(\epsilon_{1}) = -\left( \frac{\partial}{\partial\epsilon_{1}} + \frac{\partial}{\partial \tilde{\epsilon}_{d}}  \right)\left\{ \tilde{A}(\epsilon_{1})f_{0}(\epsilon_{1})    \right\}.
\label{eqAA8}
\end{eqnarray} 
Thus, inserting Eq.~(\ref{eqAA8}) into the integral $I_{1}$ we have
\begin{eqnarray}
I_{1} = -\int d\epsilon_{1} \int d\epsilon_{2} \tilde{A}(\epsilon_{2})\tilde{A}(\epsilon_{1} + \epsilon_{2} - \omega)\tilde{f}(\epsilon_{2})\tilde{f}(\omega - \epsilon_{1} - \epsilon_{2})
\left( \frac{\partial}{\partial\epsilon_{1}} + \frac{\partial}{\partial \tilde{\epsilon}_{d}}  \right)
\left\{
\tilde{A}(\epsilon_{1}) f_{0}(\epsilon_{1})\right\}.
\label{eqAA9}
\end{eqnarray}
Shifting the integration variable $\epsilon_{1}\rightarrow \epsilon_{1} + \omega$, leads to
\begin{eqnarray}
I_{1} &=& -\int d\epsilon_{1} \int d\epsilon_{2} \tilde{A}(\epsilon_{2})\tilde{A}(\epsilon_{1} + \epsilon_{2})f_{0}(\epsilon_{2})f_{0}( - \epsilon_{1} - \epsilon_{2})
\left( \frac{\partial}{\partial\omega} + \frac{\partial}{\partial \tilde{\epsilon}_{d}}  \right)
\left\{
\tilde{A}(\epsilon_{1}+\omega) f_{0}(\epsilon_{1}+\omega)\right\}\nonumber\\
&=& -\left( \frac{\partial}{\partial\omega} + \frac{\partial}{\partial \tilde{\epsilon}_{d}}  \right)
\int d\epsilon_{1} \int d\epsilon_{2} \tilde{A}(\epsilon_{2})\tilde{A}(\epsilon_{1} + \epsilon_{2})f_{0}(\epsilon_{2})f_{0}( - \epsilon_{1} - \epsilon_{2})
\tilde{A}(\epsilon_{1}+\omega) f_{0}(\epsilon_{1}+\omega)\nonumber\\
&=& -\left( \frac{\partial}{\partial\omega} + \frac{\partial}{\partial \tilde{\epsilon}_{d}}  \right)
\int d\epsilon_{1} \int d\epsilon_{2} \tilde{A}(\epsilon_{2})\tilde{A}(\epsilon_{1} + \epsilon_{2}-\omega)f_{0}(\epsilon_{2})f_{0}(\omega - \epsilon_{1} - \epsilon_{2})
\tilde{A}(\epsilon_{1}) f_{0}(\epsilon_{1}),
\label{eqAA10}
\end{eqnarray}
where, in the last line, we have shifted back the integration variable $\epsilon_{1}\rightarrow\epsilon_{1}-\omega$.

The same type of manipulations can be applied to integrals $I_{2}$ and $I_{3}$. 
Adding the three terms together, we therefore obtain
\begin{eqnarray}
\left.\frac{\partial}{\partial V}\tilde{\Sigma}^{<}(\omega,T,V)\right|_{V=0} = - \gamma \left( \frac{\partial}{\partial\omega} + \frac{\partial}{\partial \tilde{\epsilon}_{d}}  \right) \tilde{\Sigma}^{<}(\omega,T,V=0),
\label{eqAA11}
\end{eqnarray}
which proves the validity of Oguri's Ward identity explicitly for the second order diagram of Fig.~\ref{fig:A1}. A general perturbative proof can be constructed along similar lines~\cite{Oguri.01}. Nonetheless, it has been argued that the identity is only valid at $V=0$ and that the limits $\omega \rightarrow 0$ and $V\rightarrow 0$ do not commute at $T=0$ but do so at non-zero temperature~\cite{Aligia.13unpl}.\\
While there are no convincing indications that the strong-coupling fixed point at $T=0$, $\omega=0$, $V=0$ is singular, it should be clear
that an important element in the explicit proof provided above, is the fact
that  the renormalized resonance level $\tilde{\epsilon}_{d}$
is independent of voltage and temperature. If this were not the case, 
there
would be additional terms proportional to first-order-in-voltage derivatives of the type $\partial \tilde{\epsilon}_{d}/\partial{V}|_{V=0}$. As a result, it would no longer be possible to collect
all terms in such a way as to satisfy the Ward identity. 
This is exactly what happens in the 
non-equilibrium self-consistent perturbative scheme
presented in~\cite{Aligia.12}. There, the bare local renormalized spectral density of Eq.~(\ref{specdensity}) is taken to be
\begin{equation}
\tilde{A}(x)=\frac{1}{\pi}\frac{\tilde{\Delta}}{(x-\tilde{\Delta}\mbox{cot}(\pi \langle n_{d\sigma}\rangle)^2+\tilde{\Delta}^2},
\label{eqAdist}
\end{equation}
where $\langle n_{d\sigma}\rangle$ is the average local occupation number of spin projection $\sigma$ and is obtained from 
\begin{equation}
\langle n_{d\sigma}\rangle=-i\int d\omega\, \tilde{G}^{<}_{\sigma}(\omega,T,V).
\end{equation}
Notice that $\tilde{G}^{<}$ is related to $\tilde{\Sigma}^{<}$
via the identity $\tilde{G}^{<} = \tilde{G}^{A}\tilde{\Sigma}^{<} \tilde{G}^{R}$,
where the hybridization term $i\tilde{\Delta}$ is included in the self-energy components  \cite{Reply.13}.
Thus, in such a perturbative scheme, the Ward identity is violated as already pointed out in reference \cite{Reply.13}. Moreover, this violation occurs even if only the Hartree value for $\langle n_{d\sigma}\rangle$
is used in Eq.(\ref{eqAdist}).

\section{References}

\providecommand{\newblock}{}

\end{document}